\documentclass[12pt,preprint]{aastex}

\def\ltsima{$\; \buildrel < \over \sim \;$}
\def\lsim{\lower.5ex\hbox{\ltsima}}
\def\gtsima{$\; \buildrel > \over \sim \;$}
\def\gsim{\lower.5ex\hbox{\gtsima}}

\begin{document}
\title{The Multi-Band Magnification Bias for Gravitational Lenses}

\author{
J.\ Stuart B.\ Wyithe\altaffilmark{1},
Joshua N.\ Winn\altaffilmark{2}, David Rusin
}

\affil{Harvard-Smithsonian Center for Astrophysics, 60 Garden St.,
Cambridge, MA 02138}

\email{swyithe@cfa.harvard.edu; jwinn@cfa.harvard.edu; drusin@cfa.harvard.edu}

\altaffiltext{1}{Hubble Fellow}
\altaffiltext{2}{NSF Astronomy \& Astrophysics Postdoctoral Fellow}

\begin{abstract}
\noindent
We present a generalization of the concept of magnification bias for
gravitationally-lensed quasars, in which the quasars are selected by
flux in more than one wavelength band.  To illustrate the principle,
we consider the case of two-band selection, in which the fluxes in the
two bands are uncorrelated, perfectly correlated, or correlated with scatter.  
For uncorrelated fluxes, we show that the
previously-held result---that the bias is the product of the
single-band biases---is generally false.  We demonstrate some
important properties of the multi-band magnification bias using model
luminosity functions inspired by observed correlations among X-ray,
optical, infrared and radio fluxes of quasars.  In particular, the bias need
not be an increasing function of each flux, and the bias can be
extremely large for non-linear correlations.  The latter fact may
account for the high lensing rates found in some X-ray/optical and
infrared/radio selected samples.
\end{abstract}


\section{Introduction}
\label{sec:intro}

If a massive galaxy lies along the line of sight to a background
quasar, the galaxy may act as a gravitational lens, magnifying and
forming multiple images of the quasar.  Beginning with the pioneering
work of \citet{tog84}, many authors have computed the number of lenses
that should appear in well-defined samples of quasars, with
particular attention given to the dependence of this statistic on the
vacuum energy density \citep{turner90,kochanek96,helbig99,sarbu01,li02}.

These calculations must take into account not only the probability
that a massive galaxy will be aligned closely enough with a background
quasar (the lensing cross-section), but also the enhancement of the
quasar flux due to lensing (the magnification bias).  This is because
quasar samples are usually defined by observed flux in some wavelength
band, and gravitational lensing boosts the observed flux, thereby
sampling a fainter portion of the quasar luminosity function.  For
example, if intrinsically faint quasars are sufficiently more numerous
than bright quasars, then a quasar with a given observed flux is more
likely to be lensed than the cross-section alone would imply.  

More recently attention has turned toward the information about galaxy mass 
profiles that can be gleaned from lens statistics.
These statistics include lensing rates 
\citep[see, e.g.,][]{keeton01b,wyithe01,li02}, the 
ratio of four-image to two-image lenses \citep[see, e.g.,][]{rusin01b,finch02},
the image separation distribution \citep[see, e.g.,][]{kochanek01}
 and the brightness distribution of central images  
\citep[see, e.g.,][]{rusin01,keeton01,keeton02a,evens02,oguri02}. 
All of these applications of lens statistics require a good understanding of 
magnification bias. 

\citet{blr91} noted that quasar samples selected by both radio and
optical flux measurements are subject to what they called a ``double
magnification bias.''  If the radio and optical fluxes from a given
quasar are nearly independent, then quasars bright in both bands are
especially likely to be lensed\footnote{ Note that the important
property of the two bands is independence, not a large separation in
wavelength (as has since been stated in the literature;
\cite{bade97}), although of course these two properties are related.}.
By assuming that gravitational lensing produces only one possible
value of magnification, and using power-law luminosity functions for
the optical and radio bands, \citet{blr91} showed that the resulting
two-band magnification bias is the product of the bias factors
computed separately for each band.

It is timely to revisit the issue of multi-band magnification bias
with a more general approach.  With the advent of large-area sky
surveys at many wavelengths, it has become possible to define samples
of thousands of quasars by their observed fluxes in X-ray, optical,
infrared, and radio bands.  Quasars appear in large numbers in, for
example, the RASS \citep[ROSAT All-Sky Survey:][]{truemper82,voges99}
and eventually ChaMP \citep[Chandra Multi-wavelength
Project:][]{wilkes01,silverman02} at X-ray wavelengths; NVSS
\citep[NRAO-VLA Sky Survey:][]{condon98} and FIRST \citep[Faint Images
of the Radio Sky at Twenty centimeters:][]{bwh95,white97} at radio
wavelengths; 2MASS \citep[Two Micron All Sky Survey:][]{kleinmann94}
at near-infrared wavelengths; and SDSS \citep[Sloan Digital Sky
Survey:][]{york00,schneider02} at optical wavelengths.
Cross-correlation of these catalogs \citep[see,
e.g.,][]{mcmahon01,ivezic02} will become an increasingly important
source of information about quasars in general, and gravitational lens
statistics in particular.

A few lenses have already been discovered using multi-band selection
criteria, at lensing rates that are larger than the 0.2--1\% 
typical of single-band lens surveys. \citet{bade97} discovered the
gravitational lens RX~J0911.4+0551 by matching RASS sources with optical
sources from Schmidt plates. Of the $\sim40$ radio-quiet X-ray--luminous 
high-redshift quasars known, two are lensed \citep{wbb99}. A
search for very red quasars through the matching of FIRST and 2MASS has
identified two gravitational lenses out of thirteen sources
\citep{gregg02,lacy02}.  None of these projects were designed
explicitly to discover gravitational lenses, although this is a
realistic possibility for the future.

In this paper we investigate the magnification bias for quasar samples
defined by measurements in multiple wavelength bands.  After
presenting the basic formalism for $N$ bands (\S\ref{sec:mbias}), we
specialize to the case of two bands and consider some illustrative
examples.  We consider the cases in which the two fluxes are
uncorrelated (\S\ref{subsec:independent}), perfectly correlated
(\S\ref{subsec:perfectly-correlated}), and correlated with non-zero
scatter (\S\ref{subsec:correlated}).  We then use a realistic
model of the optical luminosity function for quasars to demonstrate a
few interesting properties of the multi-band magnification bias
(\S\ref{sec:examples}); in particular, the bias does not necessarily
increase with flux in each band, and there is a profound difference
between the case of a linear correlation and a non-linear correlation
with flux in another band. Finally, in \S\ref{sec:discussion} we
summarize our results, and discuss possible applications of this formalism to real
quasar samples.

\section{Magnification Bias and the Multiple Imaging Rate}
\label{sec:mbias}

We begin by reviewing the case of single-band magnification bias
\citep{turner80,tog84}.  In a sample of quasars at redshift $z$
with (apparent\footnote{By ``apparent,'' we mean that $L_1$ is the
luminosity inferred from the observed flux and the luminosity
distance, without taking into
account the possible magnification due to lensing.}) luminosity $L_1$,
the fraction of multiple-image lensed quasars is
\begin{equation}
F(L_1,z) \approx
\frac{B_1\tau_{\rm mult}}{B_1\tau_{\rm mult}+(1-\tau_{\rm mult})},
\end{equation} 
where $\tau_{\rm mult}$ is the cross-section for multiple imaging, and
$B_1(L_1)$ is the magnification bias.  For $\tau_{\rm mult}\ll1$ and 
$B_1\tau_{\rm mult}\ll1$, this reduces to the
usual expression $F(L_1,z) = B_1\tau_{\rm mult}$.
The magnification bias is evaluated as
\begin{equation}
\label{single-band-mbias}
B_1(L_1,z) = \frac{\int_0^\infty \frac{d\mu}{\mu}\frac{dP}{d\mu}\Phi_1(L_1/\mu,z)}{\Phi_1(L_1,z)},
\end{equation}
where $\Phi_1(L_1,z)$ is the quasar luminosity function, $\mu$ is the sum
of the unsigned magnifications of the multiple images, and
$\frac{dP}{d\mu}$ is the probability distribution for $\mu$, taken for
a singular isothermal sphere throughout the paper ($dP=8\mu^{-3}d\mu$ for $\mu\ge2$).  
This expression can be understood as a likelihood ratio.  The denominator
is the likelihood that the quasar is drawn from the sample of unlensed
quasars with luminosity $L_1$ (within $dL_1$).  The numerator is the
likelihood that the quasar is drawn from the fainter sample of quasars
with luminosity $L_1/\mu$ (within $dL_1/\mu$), summed over all possible
values of $\mu$.

Understood this way, the generalization to $N$ bands is
straightforward. We require knowledge of the multivariate luminosity
function, $\Phi_{\rm N}(L_1,L_2,L_3,...,L_{\rm N},z)$. For a point source, the
magnification is the same for all bands, because gravitational lensing
is achromatic\footnote{Gravitational lensing magnification is sensitive to source size.
Therefore if the emission regions for the two bands differ greatly in their spatial 
extent, then there is the possibility of wavelength-dependent magnification. For example,
the small optical emission region of a quasar may be microlensed by stars in the lensing 
galaxy, whereas the more extended emission regions at infrared or radio wavelengths
is not generally microlensed (see Wyithe \& Turner~(2002) for a recent discussion of 
microlensing and magnification bias). We ignore the possibility of microlensing in 
this paper,
but note that since the mean magnification of microlensed sources equals the 
magnification of un-microlensed sources, the results presented will be qualitatively 
correct, even for correlations involving bands that are subject to microlensing.}.
The multi-band magnification bias is therefore
\begin{equation}
\label{eq:multi-band-mbias}
B_{1...N}(L_1,L_2,L_3,...,L_N,z) =
\frac{\int_0^\infty d\mu\frac{1}{\mu^N}\frac{dP}{d\mu}
    \Phi_{1...N}(L_1/\mu,L_2/\mu,L_3/\mu,...,L_N/\mu,z)}
{\Phi_{1...N}(L_1,L_2,L_3,...,L_N,z)}.
\end{equation}

The dependence of $B$ on the apparent luminosities of the quasars depends on
the correlations, if any, between the intrinsic luminosities of the
quasars in those bands.  To illustrate the interesting properties that
can result, in the following sections we concentrate on the simplest
non-trivial case, the two-band magnification bias. All of the results
are easily generalized to $N$ bands.

\subsection{Two-Band Magnification Bias: No Correlation}
\label{subsec:independent}

\citet{blr91} considered quasars observed at both optical and radio
wavelengths, and assumed a power-law luminosity function for each
band. They showed that if the fluxes in these bands are statistically
independent, and if there is only one possible value of the lensing
magnification, then the two-band magnification bias is equal to the
product of the biases that would be computed separately for the
optical and radio bands.  This result is not true in general.  As we
show below, even if the two bands are independent, the result does not
hold because real gravitational lenses produce a distribution of
magnifications.

First, we reproduce the result of \citet{blr91} using our
formalism. For $N=2$, Eq.~(\ref{eq:multi-band-mbias}) is
\begin{equation}
\label{eq:two-band-mbias}
B_{12}(L_1,L_2,z) =
  \frac{\int_0^\infty
            \frac{d\mu}{\mu^2}\frac{dP}{d\mu}\Phi_{12}(L_1/\mu,L_2/\mu,z)}
       {\Phi_{12}(L_1,L_2,z)}.
\end{equation}
If the bands are independent, then $\Phi_{12}(L_1,L_2,z) =
\Phi_1(L_1,z)\Phi_2(L_2,z)$.
The lens model used by \citet{blr91} can be described by
$\frac{dP}{d\mu}=\delta(\mu-\mu_0)$, in which case
\begin{equation}
B_{12}(L_1,L_2,z) =
  \frac{1}{\mu_0^2}
  \frac{\Phi_{12}(L_1/\mu_0,L_2/\mu_0,z)}
       {\Phi_{12}(L_1,L_2,z)} =
  \frac{1}{\mu_0^2}
  \frac{\Phi_1(L_1/\mu_0,z)\Phi_2(L_2/\mu_0,z)}
       {\Phi_1(L_1,z)\Phi_2(L_2,z)} =
B_1(L_1,z)B_2(L_2,z).
\end{equation}

This results fails for the more realistic case in which there is a
range of possible magnifications, because $\frac{dP}{d\mu}$ appears
once in the numerator of the multi-band magnification bias, but
appears separately in each numerator in the product of the single-band
biases.  For example, following \citet{blr91},
suppose $\Phi_1(L_1,z)=\Phi_{1,*}L_1^{\alpha_1}$ and
$\Phi_2(L_2,z)=\Phi_{2,*}L_2^{\alpha_2}$.  If we adopt the
magnification distribution appropriate for an isothermal sphere
($\frac{dP}{d\mu}=\frac{8}{\mu^3}$ for all $\mu\ge2$), then
\begin{equation}
\label{eq:two-band-mbias-power-law}
B_{12}(L_1,L_2,z) =
\int_2^\infty  \frac{d\mu}{\mu^2}\frac{8}{\mu^3}
               \frac{1}{\mu^{\alpha_1}\mu^{\alpha_2}} =
\frac{8}{4+\alpha_1+\alpha_2} 2^{-(4+\alpha_1+\alpha_2)},
\end{equation}
for $\alpha_1+\alpha_2>-4$.
Analogous calculations of the single-band bias factors give
\begin{equation}
\label{eq:single-band-biases-power-law}
B_1(L_1,z)B_2(L_2,z) =
\frac{16}{(3+\alpha_1)(3+\alpha_2)}2^{-(4+\alpha_1+\alpha_2)},
\end{equation}
which can be either larger or smaller than Eq.~(\ref{eq:two-band-mbias-power-law}).

\subsection{Two-Band Magnification Bias: Perfect Correlation}
\label{subsec:perfectly-correlated}

If the two bands are perfectly correlated, with $L_2 = f(L_1)$, one might
expect that no new information is provided by the observation in the
second band, and therefore that the two-band bias is equal to the
single-band bias for either band. This is not quite true.  Gravitational 
magnification
multiplies both fluxes by the same factor.  If the unmagnified fluxes
are linearly correlated, then the magnified fluxes also obey the
correlation.  However, if the correlation is non-linear, then the
magnified fluxes do not obey the correlation, and the source must be
gravitationally lensed. In the the appendix, we derive this result formally, 
by calculating the magnification bias for general correlations 
(see the next section) in the limit of zero scatter.

\subsection{Two-Band Magnification Bias: Imperfect Correlation}
\label{subsec:correlated}

\begin{figure*}[htbp]
\epsscale{0.5}
\plotone{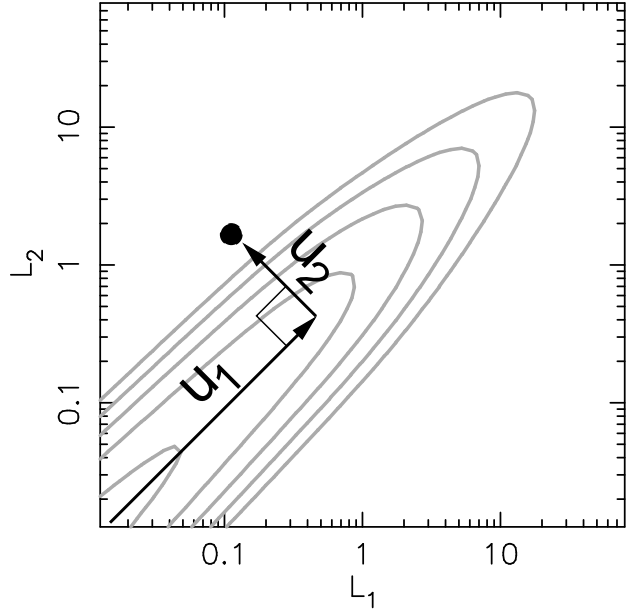}
\caption{
\label{fig0}
Schematic of variables defining the luminosity function for non-zero scatter 
about a power-law correlation. The grey lines are contours of a bi-variate 
luminosity function and are shown for context.}
\end{figure*}

More generally, $L_1$ and $L_2$ are correlated with some intrinsic
scatter. On physical grounds we expect the magnitude of the scatter 
to scale with the luminosity [i.e. $\Delta L_2/L_2\sim g(\Delta L_1/L_1)$],
which makes it convenient to use logarithmic variables $l=\log L$.
Suppose that the correlation between $L_1$ and $L_2$ is a power-law, $L_1=L_2^\gamma$,
or $l_1=\gamma l_2$.
Because of the correlation, it is convenient to express the luminosity function 
in terms of the new variables 
$u_1\equiv\frac{1}{\gamma} l_1+l_2$ and $u_2\equiv-\gamma l_1+l_2$, 
which describe the location parallel and perpendicular to the correlation, 
respectively (see Fig.~\ref{fig0}). In these variables, the luminosity function
(expressed in density per square logarithmic interval) is
$\Psi_{12}(u_1,u_2,z)=(\gamma+\frac{1}{\gamma})\Phi_{12}(l_1,l_2,z)$. The luminosity 
function can also be written $\Psi_{12}(u_1,u_2,z)=\Psi_1(u_1,z)p(u_2|u_1,z)$, 
where $\Psi_{1}(u_1,z)$ is the luminosity function in the new variable $u_1$ and 
$p(u_2|u_1,z)$ is the conditional probability of $u_2$ given $u_1$. Because we expect
the scatter to be symmetric in reflection about the correlation\footnote{As an 
example of why we expect the scatter to be symmetric in reflection about the 
correlation, consider a sample of quasars 
with flux measured in two optical wave bands, say $r$ and $i$. We would expect 
to find a variation of bias with $i$-band at fixed $r$-band that is 
qualitatively similar to the variation with $r$-band at fixed $i$-band. 
This symmetry in the magnification bias requires symmetry 
of the scatter in reflection about the correlation. We therefore choose 
to model the scatter as a symmetric function in logarithms of luminosity;
that is, defined normal to the correlation (i.e. along the $u_2$ axis).}
, we assume
 $p(u_2|u_1,z)$ to be Gaussian with variance $\sigma$, hence
\begin{equation}
\label{transform_bias}
\Psi_{12}(u_1,u_2,z)=\Psi_{1}(u_1,z)\frac{1}{\sqrt{2\pi}\sigma}\exp{ \left(-\frac{u_2^2}{2\sigma^2}\right)}.
\end{equation}
Defining $M=\log\mu$, the magnification bias is
\begin{eqnarray}
\nonumber
B_{12}(l_1,l_2,z)&=&\int_0^\infty dM \frac{dP}{dM} \frac{\Phi_{12}(l_1-M,l_2-M,z)}{\Phi_{12}(l_1,l_2,z)}\\
&=&\int_0^\infty dM \frac{dP}{dM} \frac{(\gamma+\frac{1}{\gamma})\Psi_{12}\left[u_1-(1+\frac{1}{\gamma})M,u_2+(\gamma-1)M,z\right]}{(\gamma+\frac{1}{\gamma})	\Psi_{12}(u_1,u_2,z)}.
\end{eqnarray}
Inserting Eq.~(\ref{transform_bias}),
\begin{equation}
\label{transform_bias2}
B_{12}(l_1,l_2,z)=\int_0^\infty dM' \frac{dP}{dM'} \frac{\Psi_{1}\left[u_1-M',z\right]}{\Psi_{1}(u_1,z)}\exp{\left(-\frac{1}{2\sigma^2}\left[\left(u_2+\frac{\gamma (\gamma-1)}{1+\gamma}M'\right)^2-u_2^2\right]\right)},
\end{equation}
where we have defined $M'=(1+\frac{1}{\gamma})M$.
This expression illustrates many important points. First, if the correlation is 
linear ($\gamma=1$) then $B_{12}$ is independent of $u_2$, and the contours of
constant bias run normal to the correlation (i.e. along lines of constant $l_2+l_1$).
Furthermore, if $\Psi_1(u_1,z)$ is a monotonically decreasing function of $u_1$, then we 
find that $B_{12}(l_1,l_2,z)$ is an increasing function of both $l_1$ and $l_2$ (since 
$u_2$ increases monotonically with both $l_1$ and $l_2$).
This example will be further explored in Case 1 of \S\ref{sec:examples}.
Eq.~(\ref{transform_bias2}) also demonstrates the behavior arising
from non-linear correlations ($\gamma\ne1$). Here the exponential plays an
important role; it introduces an asymmetry in the bias across
 the correlation. If, for example, $\gamma>1$, then large biases can result from 
negative values of $u_2$ (i.e. below the correlation), because the exponent becomes 
positive. On the other hand, the exponent is negative for all $u_2>0$, and hence the bias 
above the correlation is small. This example will be further explored in Case 3 of 
\S\ref{sec:examples}.

In the following section, we evaluate Eq.~\ref{transform_bias2} numerically, with more
realistic assumptions, in order to illustrate these and  other interesting
and potentially observable properties of the multi-band magnification
bias.

\section{Magnification Bias for Illustrative Bi-Variate Luminosity Functions} 
\label{sec:examples}

We consider measurements made in two bands, and a power-law
correlation between the two bands:
$L_2=L_1^\gamma$.  As in \S\ref{subsec:correlated}, the scatter (normal to the 
correlation) is assumed to be Gaussian in logarithms with half-width 
$\sigma$.  We consider 4 examples of
Eq.~(\ref{transform_bias}).  The first two examples involve linear
correlations ($\gamma=1$), and the second two examples involve
non-linear correlations.

For the first of the two correlated bands, we use a luminosity function
$\Phi_1(L_1,z)$ that is appropriate for optical wavelengths.
A good representation of the observed optical quasar luminosity
function at redshifts $z\la 3$ is provided by the following double
power-law form \citep{bsp88,pei95}:
\begin{equation}
\label{LF}
\Phi_{\rm o}(L,z) = 
\frac{\Phi_*/L_{*}(z)}
     {[L/L_{*}(z)]^{\beta_{\rm l}}+[L/L_{*}(z)]^{\beta_{\rm h}}}.
\end{equation}
At the faint end, the logarithmic slope of this function is
$-\beta_{\rm l}=-1.58$, while at the bright end the slope 
is $-\beta_{\rm h}=-3.43$ (Boyle et al.~2000).
Moreover, all dependence on redshift (for $z\la3$) is in the break
luminosity $L_{*}(z)$.  We therefore show the luminosity function in
units of $L_{*}$ throughout the remainder of this paper. 
Setting $\Phi_1(L_1)=\Phi_{\rm o}(L_1)$, we find from 
Eq.~\ref{transform_bias} that $\Phi_{\rm o}(L_1)$ is related to 
$\Psi_{1}(u_1,u_2)$ through
\begin{equation}
\Phi_{\rm o}(L_1)=\int_0^{\infty}dL_2\Psi_{1}(u_1)\frac{1}{\sqrt{2\pi}\sigma}\exp{ \left(-\frac{u_2^2}{2\sigma^2}\right)}.
\end{equation}
This equation defines the functions $\Psi_1(u_1)$ used in this section.

\begin{figure*}[htbp]
\epsscale{1}
\plotone{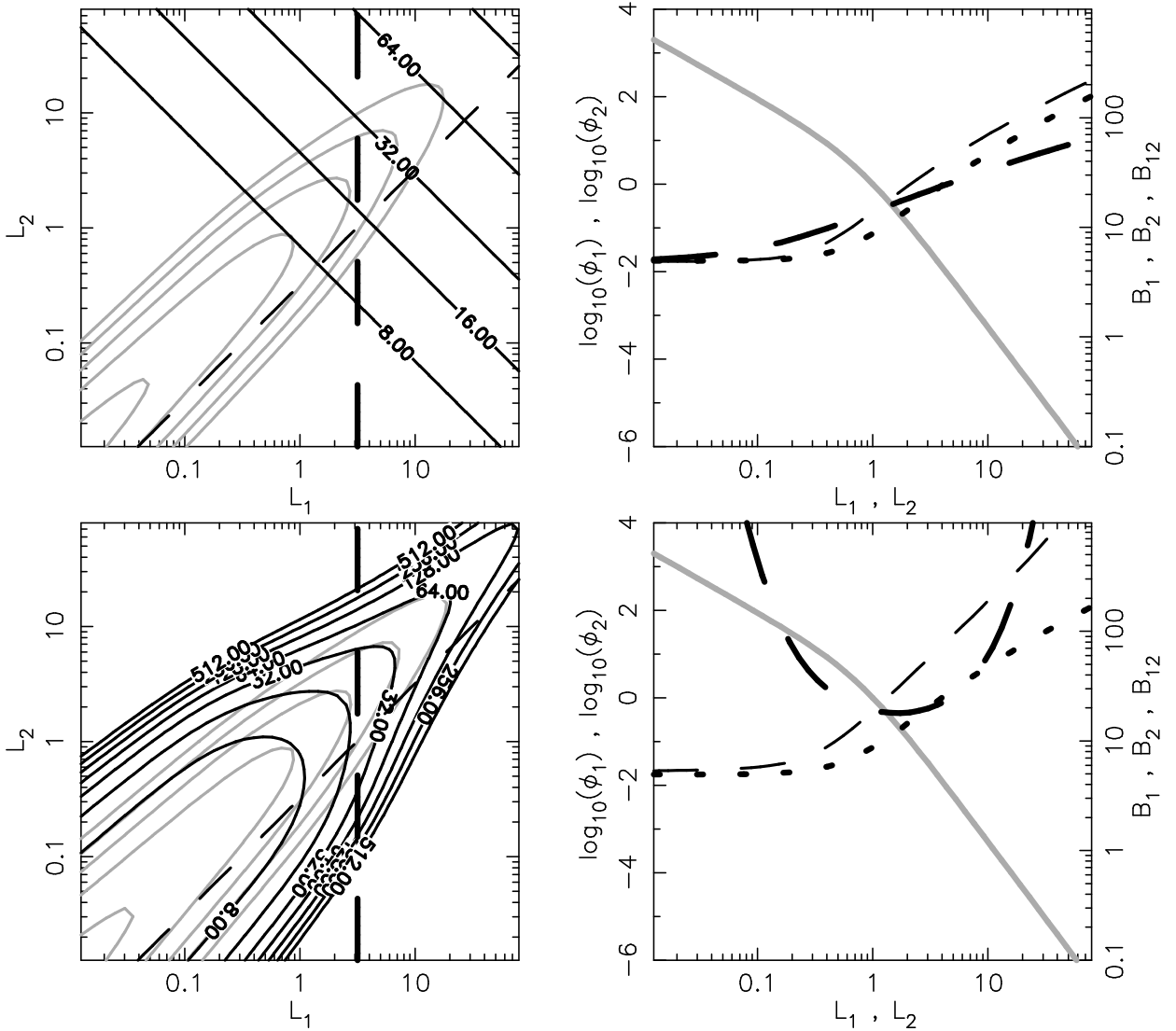}
\caption{
\label{fig:bias-linear-correlation}
Bi-variate magnification biases for linear correlations with
scatter. Top: Linear correlation, with a scatter that is insensitive to 
luminosity ($\gamma=1.0$, $\sigma=0.15$). Bottom: Linear correlation with a scatter
that decreases with luminosity ($\gamma=1.0$, $\sigma=0.15-0.02\log u_1$).
The left hand panels show contours of
the bi-variate luminosity function (grey lines).  The solid 
lines are contours of
magnification bias. The right hand panels show the corresponding
single band luminosity functions (grey lines). Also shown are the
single band magnification biases (dotted lines), and the
magnification bias along the paths denoted by the dashed lines in the
left hand figure. The bias for the path denoted by the thin dashed line is 
plotted as a function of $L_{\rm 1}$, while the bias along the thick dashed
line is plotted as a function of $L_{2}$. Because of the linear 
correlation, the single band luminosity functions and magnification 
biases are identical for the two bands.}
\end{figure*}

\begin{enumerate}

\item $\gamma=1.0$, $\sigma=0.15$: A linear correlation with
constant scatter.  This is the situation that might be expected
between two different optical bands. The contours of $\Phi_{12}$
(gray), and of $B_{12}$ (black), are shown in the top left panel of
Fig.~\ref{fig:bias-linear-correlation}.  As was derived in the previous 
section (and might be expected intuitively), the bias increases 
monotonically with both luminosities, and is constant along lines
normal to the correlation.

The top right panel shows the corresponding single-band luminosity
function (gray line), and magnification bias (dotted black line).
These functions are the same for both bands because of the linear
correlation. In addition we plot $B_{12}(L_1,L_2)$ for two paths
through $(L_1,L_2)$-space: one at fixed $L_1$ (thick dashed line, 
plotted as a function of $L_2$) and the other below but parallel to the 
correlation (thin dashed line, plotted as a function of $L_1$).

\item $\gamma=1.0$, $\sigma=0.15 - 0.02 \log u_1$: 
Same as the previous
example, except in this case we allow the logarithmic scatter to depend on
luminosity.  The results are shown in the lower panels of
Fig.~\ref{fig:bias-linear-correlation}, in the same format as the
previous example.

The contours of magnification bias wrap around the contours of
$\Phi_{12}$, increasing rapidly as one moves normal to the correlation
(along the $u_2$ axis).  This can be understood as follows.
Magnification draws quasars from regions of lower intrinsic
luminosity, where the probability density of quasars (as given by
$\Phi_{12}$) is larger.  This is especially true when the observed 
luminosities fall at some distance from the
correlation, because the scatter in the correlation is larger at lower
luminosities.

The dependence of $B_{12}$ on $L_2$, for fixed $L_1$, is again shown
by the thick dark dashed line. Interestingly, the dependence is not
monotonic. For small values of $L_2$ the magnification bias is large.
The bias decreases as $L_2$ rises through the expected intrinsic
value, and then increases again. The bias along the path denoted by the 
thin dashed line demonstrates that the bias can become very large 
for sources below the correlation.

\item $\gamma=1.5$, $\sigma=0.2$: A non-linear correlation with a
constant scatter.  This situation approximates the correlation that
has been observed between the X-ray ($L_2$) and optical ($L_1$) bands for quasars
\citep{brinkmann00}.  Results for this case are shown in the top two
panels of Fig.~\ref{fig:bias-non-linear-correlation}, in the same
format as the previous examples.

In this case, $\Phi_2(L_2)$ (thin gray line) is a flatter function
than $\Phi_1(L_1)$ (thick gray line), and therefore the single-band bias
$B_2$ (thin dotted line) is smaller than $B_1$ (thick dotted line).  
Although $B_{12}$ is an increasing
function of $L_1$, it is actually a decreasing function of $L_2$ (for fixed
$L_1$). This runs counter to the naive expectation that the brighter the
quasar is (regardless of band), the more likely it is to be lensed.
The reason is that when $L_2$ is smaller than expected from the
correlation, reducing both luminosities by the same factor $\mu$
(along a line of unit slope, in the top left panel) brings one to a
region of much higher probability.

\item $\gamma=-1.0$, $\sigma=0.3$: An anti-correlation with constant
scatter.  This is a somewhat artificial example, but we might imagine
there are two ways for a quasar with a fixed energy source to radiate
its energy, and one of these ways can be blocked by a variable amount.
For example, the optical and far-infrared luminosities might be
expected to exhibit some degree of anti-correlation due to dust
obscuration.

The results are plotted in the lower panels of
Fig.~\ref{fig:bias-non-linear-correlation}.  The contours of
magnification bias are parallel to the anti-correlation. For small
luminosities the bias is smaller than unity. Quasars in this region
are {\em less} likely to be lensed than the cross-section alone would
imply, because lensed quasars would be drawn from a population with
very small density.  Conversely, for large luminosities, the bias
becomes arbitrarily large.

\end{enumerate}

\section{Discussion}
\label{sec:discussion}

The multi-band magnification bias is an {\em a posteriori}
statistic. It is used to estimate the probability that the apparent
luminosities of a given quasar, as measured in several bands, are due to
gravitational magnification, rather than being intrinsic to the
quasar.  When a sample of quasars is selected through the matching of
sources in two different catalogs, both fluxes must be used to perform
this calculation.  One must also have some knowledge of the intrinsic
correlation (if any) of the fluxes, and the distribution of
magnifications produced by lensing.

One might expect that the multi-band bias is maximized when the bands
are uncorrelated (an example of which is shown for the radio-optical 
correlation of SDSS early data release quasars in Fig.~\ref{fig3}), 
since in that case there is no redundant information
in the flux measurements.  Upon further reflection, or using the
mathematics developed in this paper, one realizes that this is not
true---the relevant information is how discrepant the observed
fluxes are from the correlation, and whether the discrepancy can be made
smaller if the observed fluxes are all reduced by a constant factor.

Many of the illustrative examples presented in this paper 
approximate certain correlations that have been observed for real
quasars.  In particular, the multi-band magnification bias may result in
very high lens fractions for certain quasar samples.  First, we
consider the case of a quasar sample selected by optical colors.  The
top left panel of Fig.~\ref{fig3} is a logarithmic plot of SDSS $i$-band {\em vs.}
$r$-band fluxes for the SDSS early data release quasars (Schneider
et al.~2002).  The data show a linear correlation with scatter, the
magnification bias for which is illustrated in the top panels of 
Fig.~\ref{fig:bias-linear-correlation}.
The magnification bias must be computed using both optical
measurements, unless the sample is 100\% complete in one filter (i.e.,
unless after selecting quasars in $i$, the $r$-band magnitude was measured in
every single case).  As an example consider the sample of SDSS $z>5.8$
quasars (Fan et al.~2001). Since the $z$-band selection is at
$\sim1100\AA$ in the rest-frame, quasars with fixed absolute B
magnitude ($\sim4400\AA$) are more likely to be selected if they are
bluer than average. Thus a sample of quasars selected in this manner will be 
bluer than average and lie blueward of the correlation on a plot of the 
intrinsic correlation between $M_{\rm V}$ and 
$M_{\rm B}$. The magnification bias for these sources may be
significantly smaller than that computed using only extrapolations of the 
$B$-band luminosity function \citep{wl02,chs02}.  

\begin{figure*}[htbp]
\epsscale{1}
\plotone{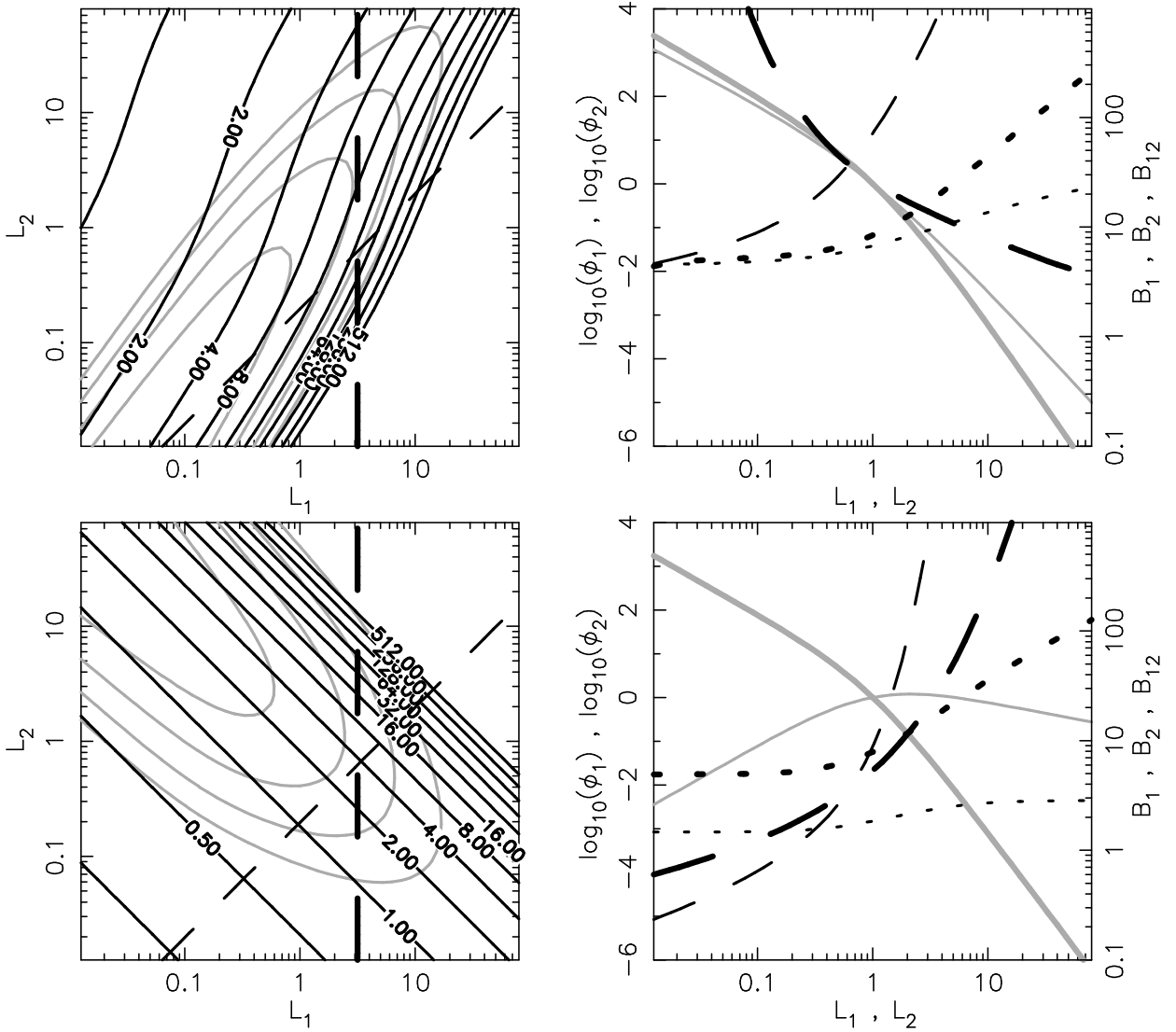}
\caption{
\label{fig:bias-non-linear-correlation}
Bi-variate magnification biases for non-linear
correlations with scatter. Top: Non-linear
correlation, with a logarithmic scatter that is insensitive to luminosity 
($\gamma=1.5$, $\sigma=0.2$). Bottom:
Non-linear anti-correlation, with a logarithmic
scatter that is insensitive to luminosity ($\gamma=-1.0$, $\sigma=0.3$). 
The left hand panels show
contours of the bi-variate luminosity function (grey lines).  The 
solid lines are
contours of magnification bias. The right hand panels show the
corresponding single band luminosity functions (grey lines) and the
single band magnification biases (dotted lines). Thick and thin
lines denote quantities in $L_{\rm 1}$ and $L_{\rm 2}$
respectively. Also shown are the magnification biases along the paths
denoted by the dashed lines in the left hand figure.  The bias for the 
path denoted by the thin dashed line is 
plotted as a function of $L_{\rm 1}$, while the bias along the thick dashed 
line is plotted as a function of $L_{2}$.}
\end{figure*}

\begin{figure*}[htbp]
\epsscale{1}
\plotone{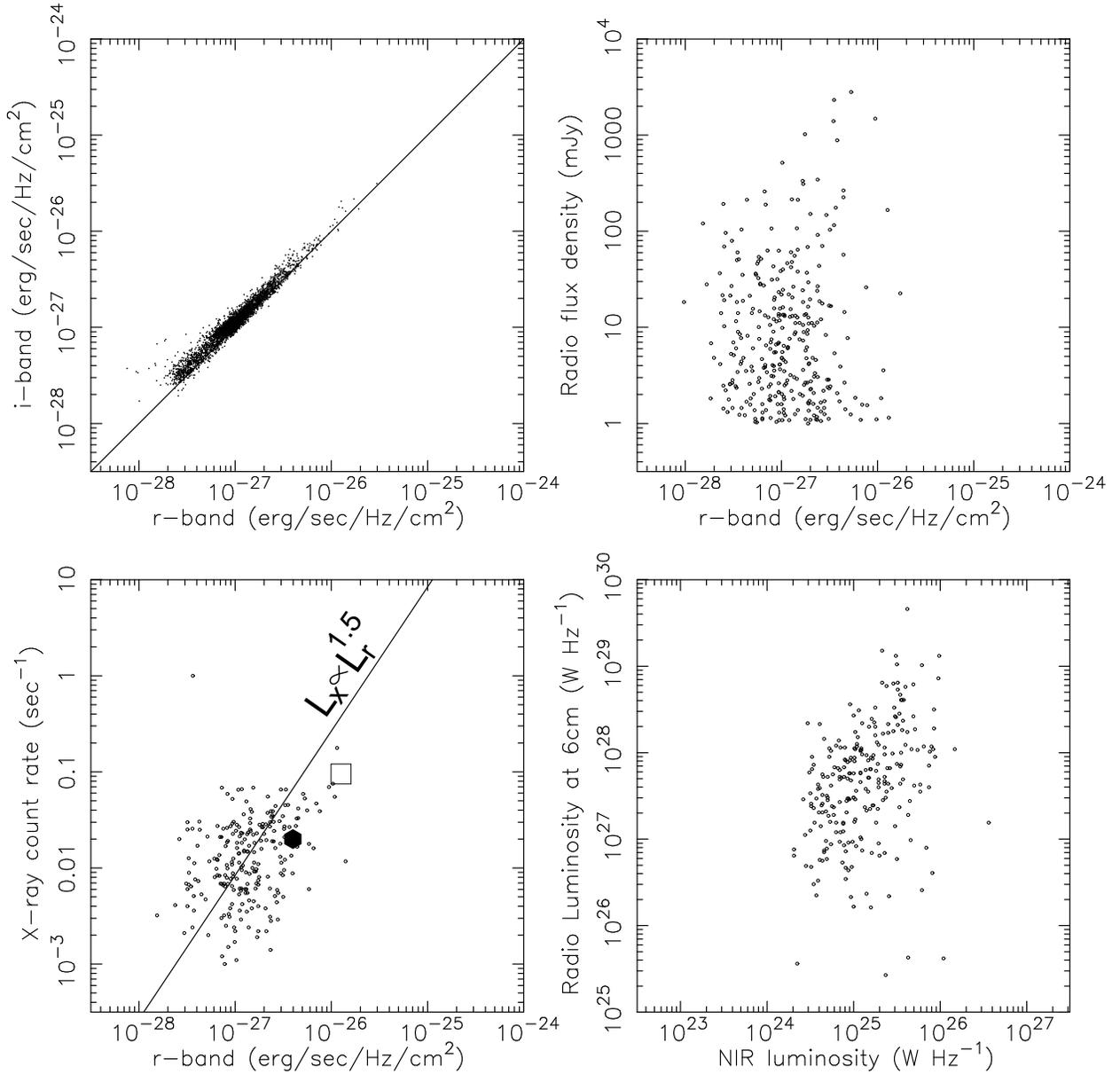}
\caption{\label{fig3} Correlations in different bands. 
Top Left: SDSS $i$-band vs. SDSS $r$-band flux. (Schneider et al.~2002) 
Top Right: FIRST radio flux vs. SDSS $r$-band flux (Schneider et al.~2002). Lower
Left: ROSAT X-ray counts vs. SDSS $r$-band flux (Schneider et al.~2002) for quasars
with redshifts larger than 0.5. 
The large dot in this panel represents RX~J0911.4+0551, while the open square 
shows the location of HE~1104-1805. Bottom Right: Radio vs. Near IR luminosity
(Barkhouse \& Hall~2001). In the left hand panels the observed correlation lines 
are drawn to guide the eye.}
\end{figure*}

Next, we consider examples of non-linear luminosity
correlations. \citet{brinkmann00} measured the correlation between
ROSAT X-ray and FIRST radio fluxes for matched quasar samples. They
find that while radio-quiet quasars show a linear relationship between
X-ray and radio luminosity, radio-loud quasars have an X-ray flux
$L_x$ that varies with the radio luminosity $L_{\rm r}$ as $L_x\propto
L_{\rm r}^{0.48\pm0.05}$ with an intrinsic scatter of $\sim0.2$
dex. Furthermore, \citet{brinkmann00} showed that the X-ray luminosity
correlates non-linearly with optical luminosity $L_{\rm o}$, following
$L_x\propto L_{\rm o}^{1.42\pm0.09}$. The second of these correlations 
(in flux) is plotted in Fig.~\ref{fig3} for quasars in the SDSS Early Data
Release \citep{schneider02}, but can be seen more clearly in
Fig.~14 of \citet{brinkmann00}. The multi-band magnification bias
corresponding to the second correlation\footnote{Note that while we have presented 
results for a non-linear correlation with index $\gamma>1$, the result for $\gamma<1$
is simply obtained through reversal of the axes.} is illustrated in the top
panels of Fig.~\ref{fig:bias-non-linear-correlation}. The magnification
bias can be extremely large for sources that are luminous at both 
optical and X-ray wavelengths. This may be the explanation for the apparently
high probability of lensing in bright X-ray selected quasar catalogs
\citep{bade97}. The location of the gravitational lens 
RX~J0911.4+0551, which was selected from cross-correlation of optical and
X-ray catalogs, is shown on this plot by the large dot\footnote{We used the integrated 
$R$ from Bade et al.~(1997) and color transformations from Fukugita, Shimasaku 
\& Ichikawa~(1995).}. The fluxes place the quasar below the correlation, in the region 
where we expect the magnification bias to be large (see the upper panels of
Fig.~\ref{fig:bias-non-linear-correlation}). The lens HE 1104--1805 is also X-ray 
loud (Wisotzki et al.~1993; Reimers et al.~1995). While this
lensed quasar was not discovered through cross-correlation between catalogs, it 
is interesting to note its location on this plot, shown 
by the open square in Fig.~\ref{fig3}. 
The quasar is found to be very bright in both bands, and is again in the region of high
magnification bias. It is suggestive that the two X-ray loud gravitational 
lenses both appear to lie in the region of high magnification bias, as expected.

Finally, the multi-band magnification bias may
also provide an explanation for the large gravitational lens fraction
(2 out of 13) found through the matching of FIRST and 2MASS sources
\citep{gregg02,lacy02}. Fig.~\ref{fig3} shows the correlation for near
infrared luminosities verses radio luminosities compiled
from table~1 of \citet{bh01}. The radio/near-IR correlation appears to
be steeper than linear. If true we might expect very large biases for
luminous near-IR sources. The top panel of Fig.~\ref{fig:bias-non-linear-correlation}
demonstrates magnification bias for a non-linear correlation, and shows that the 
bias of around 100 necessary to achieve a lens fraction of 2/13 is possible.

\section{Summary}
\label{sec:summary}

This paper has discussed the multi-band magnification bias for
gravitational lensing with arbitrary luminosity functions in several
bands.  Previous discussion of the
multi-band magnification bias \citep{blr91} focused on the case where
the fluxes in the two bands are independent.  If a single value for
the lens magnification is considered, they showed that this assumption
leads to a multiple magnification bias that is equal to the product of
the single-band biases.  However, we have shown that this equality
breaks down in the more realistic case when there is a
distribution of possible magnifications.

We also discussed the multi-band magnification bias when the fluxes in
the various bands are correlated.  In the case of a perfect (i.e.\ zero scatter)
linear correlation, the information from the second band does not
change the magnification bias. However, if the correlation is 
non-linear, then sources with
fluxes that obey the correlation cannot be lensed. On the other hand, sources
with fluxes that do not obey the correlation must be lensed.

Of course, real correlations have intrinsic scatter. We have calculated the
multi-band magnification bias for bi-variate luminosity functions with
finite scatter about both linear and non-linear correlations.  For a
linear correlation (as expected for a quasar sample selected by
optical colors) we find that the magnification bias is an increasing
function of either flux.  Calculations of lens statistics from
incomplete color-selected quasar samples should therefore account for the
multi-band magnification bias.

Non-linear correlations (and anti-correlations) with finite scatter
were also explored. If the fluxes in two bands are correlated through
a relation that is steeper than linear, then sources that lie below the 
correlation can be subject to a very large bias. The observed
correlation between X-ray and optical flux (and possibly between
infrared and radio flux) for quasars is steeper than linear. Suggestively, the two known 
X-ray loud gravitationally lensed quasars lie below the X-ray/optical correlation
in the region of large magnification bias. Thus the
multiple magnification bias may provide an explanation for the large
lensing rates found in X-ray/optical and infrared/radio selected
samples.

\acknowledgements The authors wish to acknowledge discussions with
Chris Kochanek that led us to work on this topic. We also thank Wai-Hong Tham
and Lara Winn for enduring conversations. J.S.B.W.\ is
supported by a Hubble Fellowship grant from the Space Telescope
Science Institute, which is operated by the Association of
Universities for Research in Astronomy, Inc., under NASA contract
NAS~5-26555. J.N.W.\ is supported by an Astronomy \& Astrophysics
Postdoctoral Fellowship, under NSF grant AST-0104347.

\newpage

\begin{appendix}

\section{Two-Band Magnification Bias for Perfect Correlations}

In this appendix we derive the results mentioned in \S~\ref{subsec:perfectly-correlated}
for the two-band magnification bias in the case of perfect correlations by taking the 
limit of small $\sigma$ in Eq.~(\ref{transform_bias2}). It is convenient to rewrite the 
exponential as follows
\begin{equation}
\label{bias_limit}
\lim_{\sigma\rightarrow0}B_{12}(l_1,l_2,z)=\int_0^\infty dM' \frac{dP}{dM'} \frac{\Psi_{1}\left[u_1-M',z\right]}{\Psi_{1}(u_1,z)}\lim_{\sigma\rightarrow0}\exp{\left(-\frac{1}{2\sigma^2}M'\frac{\gamma(\gamma-1)}{\gamma+1}\left[\frac{\gamma (\gamma-1)}{\gamma+1}M'+2u_2\right]\right)}.
\end{equation}

First, consider the case of a linear correlation ($\gamma=1$). In this case we find the 
exponential function is unity for all $\sigma>0$, and Eq.~(\ref{bias_limit}) reduces to  
\begin{equation}
\lim_{\sigma\rightarrow0}B_{12}(l_1,l_2,z)=\int_0^\infty dM' \frac{dP}{dM'} \frac{\Psi_{1}\left[u_1-M',z\right]}{\Psi_{1}(u_1,z)} = \int_0^\infty dM \frac{dP}{dM} \frac{\Psi_{1}\left[2(l_1-M),z\right]}{\Psi_{1}(2l_1,z)}.
\end{equation}
Note that in this case we must have $l_1=l_2$. We can  
relate $\Psi_{1}(u_1,z)$ to $\Phi_{1}(l_1,z)$ in the limit of small $\sigma$:
\begin{eqnarray}
\nonumber
\lim_{\sigma\rightarrow0}\Phi_{1}(l_1,z)&=&\int_0^\infty dl_2\lim_{\sigma\rightarrow0}\Phi_{12}(l_1,l_2)=\int_0^\infty dl_2\lim_{\sigma\rightarrow0}\frac{\Psi_{12}(u_1,u_2)}{2}=\frac{1}{2}\int_0^\infty dl_2\Psi_1(u_1)\delta(u_2)\\&=&\frac{1}{2}\int_0^\infty dl_2\Psi_1(l_1+l_2)\delta(l_1+l_2)=\frac{1}{2}\Psi_1(2l_1).
\end{eqnarray}
Thus, for a perfect linear correlation, we find that
\begin{equation}
\lim_{\sigma\rightarrow0}B_{12}(l_1,l_2,z)=\int_0^\infty dM \frac{dP}{dM} \frac{\lim_{\sigma\rightarrow0}\Phi_{1}\left[l_1-M,z\right]}{\lim_{\sigma\rightarrow0}\Phi_{1}(l_1,z)}=B_{1}(l_1,z),
\end{equation}
which is the single band magnification bias. Therefore, if two luminosities obey a perfect
linear correlation, the magnification bias is simply equal to the single-band
magnification bias computed from either luminosity.

Next consider the case of a non-linear correlation ($\gamma\ne1$). 
Specifically, we assume $\gamma>1$ and $0<M'_{\rm min}<M'<\infty$. Then
\begin{eqnarray}
\nonumber
\lim_{\sigma\rightarrow0}\exp{\left(-\frac{1}{2\sigma^2}M'\frac{\gamma(\gamma-1)}{\gamma+1}\left[\frac{\gamma (\gamma-1)}{\gamma+1}M'+2u_2\right]\right)}&=& 0 \hspace{10mm}\mbox{if}\hspace{5mm} u_2>-\frac{\gamma(\gamma-1)}{\gamma+1}\frac{M'}{2}\\
\nonumber
                                                              &=& 1 \hspace{10mm}\mbox{if}\hspace{5mm} u_2=-\frac{\gamma(\gamma-1)}{\gamma+1}\frac{M'}{2}\\
                                                              &=& \infty \hspace{10mm}\mbox{if}\hspace{5mm} u_2<-\frac{\gamma(\gamma-1)}{\gamma+1}\frac{M'}{2}.
\end{eqnarray}
Note that only sources with $u_2=0$ or 
$u_2<-\frac{\gamma(\gamma-1)}{\gamma+1}M'_{\rm min}$ are allowed.
As a  result we find that $\lim_{\sigma\rightarrow0}B_{12}(l_1,l_2,z)=\infty$ if 
$u_2<-\frac{\gamma(\gamma-1)}{\gamma+1}M'_{\rm min}$. Sources that lie on the correlation 
have $u_2=0$, and therefore $\lim_{\sigma\rightarrow0}B_{12}(l_1,l_2,z)=0$. Thus if the 
correlation is non-linear, sources that lie on the correlation cannot be lensed, while 
sources that lie off the correlation must be lensed.

\end{appendix}

\newpage

\end{document}